\documentstyle[12pt]{article}
\begin{document}
\begin{center}
{\large {\bf Morita Equivalence and Interpolation of The Dirac-Born-infeld 
Theory on the Non-Commutative Torus
}}\\
\vspace{1cm}
Pei Wang ~~~~~~Rui-Hong Yue ~~~~~~Kang-Jie Shi\\
\vspace{5mm}
 {\sf Institute of Modern Physics, Northwest University,
Xi'an, 710069,  China}
\end{center}
\footnote {e-mail: pwang@phy.
nwu.edu.cn,\quad yue@phy.nwu.edu.cn,\quad kjshi@phy.nwu.edu.cn.}
\vspace{2cm}  
\begin{abstract}
In the noncommutative Dirac-Born-Infeld action with Chern-Simons term,
an interpolation field $\Phi$ is used in both DBI action and Chern-Simons
term. The Morita equivalence is discussed in both the lagrangian  and the
Hamiltonian formalisms, which is more transparent in this treatment.
 \end{abstract}  
 
\medskip

\medskip

\medskip
\par
In recent two years, the application of noncommutative geometry in
string/M theory has got great development. This started with a paper
for describing M-theory compactifying on a noncommutative two-torus [1].
 Following it there appeared a lot of papers, some emphasizing matrix
model [2], others emphasizing D-brane [3] (a more complete list can be 
found from [4]). In these researches, a kind of new symmetry called Morita
equivalance has been studied extensively [5-10]. The correlation with the 
T-duality of type II string is also elaborated [8,9,11].

 In reference [4], Seiberg and Witten derived noncommutative
Yang-Mills theory from the quantization of open string, ending on
D-brane
in the presence of a NS-NS B-field. They argued
that whether the commutativity or noncommutativity of Yang-Mills theory 
depends on the choice of regularization and proved their equivalence
through the DBI action. Especially they proposed that there is an
interpolating theory between these two  with a modulus $\Phi$ which can
be considered as a magnetic background [6,11].

 Using this interpolating scheme Ryang examined the Morita equivalence
of DBI action [10]. He got same Morita transformation rule for the
noncommutative open string parameters as reference [4] but without
recourse to the low
energy zero slope limit $\alpha' \rightarrow 0$. His proof of the
Morita transformation invariance is not only directly in DBI Lagrangian
form, but also much simpler than other methods (for example Ref. [9]).
However, he didn't let the interpolation cover the whole action, which 
includes a Chern-Simons topological term.  The $C$ fields in this term
change in a somehow complicated way under Morita transformation.
Because of this reason we would
like to suggest the interpolating with modulus $\Phi$ in both
the DBI action and the Chern-Simons term.

 Morita equivalence is related to the T-duality of type-II string in
which D-branes are compactified on a p-torus. So we need to consider the 
Dirac-Born-Infeld action. For the convenience of comparision we use the
same notations as references [4,10]. For ordinary D-brane we have 
\begin{eqnarray}
S &=& \int{\rm d}^{p+1}\sigma{\cal L}\quad,\quad  {\cal L}={\cal L}_{DBI}+
{\cal L}_{wz},\\
{\cal L}_{DBI}&=& -\frac1{g_s(2\pi)^p(\alpha')^{\frac{p+1}{2}}}
STr\sqrt{-det_{(p+1)}(g+2\pi\alpha' (F+B))},
\end{eqnarray}
in which $g_{s}, g, B$ belong to closed string parameters, and we use the
symmetric trace for non-Abelian gauge group [12]. 
\begin{eqnarray}
{\cal L}_{wz}&=& STr(e^{2\pi \alpha' (F+B)} \sum C_{(n)})=STr
P_{(p)}(C,F+B),
\end{eqnarray}
where $P_{(p)}(C,F+B)$ is a polynomial of R-R potentials and the second
equality is valid under the integral of world volume of D-p branes. But
we are
interested in the noncommutative D-brane with modulus $\Phi$. Thus the 
Lagrangian is 
\begin{eqnarray}
\hat{{\cal L}}_{DBI}&=& -\frac{1}{G_{s}(2\pi)^{p}\alpha' \;
^{\frac{p+1}{2}}}
STr_{\theta} \sqrt{-det_{(p+1)}(\hat {G}+{\cal F})},\\ 
 \hat{{\cal L}}_{wz}&=&STr_{\theta} P_{(p)}(\hat{C},{\cal F}),\quad\quad
\hat{{\cal L}}=\hat{{\cal L}}_{DBI}+\hat{{\cal L}}_{wz}=Str_{\theta}L.
\end{eqnarray}
in which ${\cal F}\equiv 2\pi{\alpha'}(\hat{F}+\Phi), G_{s},
\hat{G}$ belong to open string and $\det_{(p+1)}$ stands for the
determinent of $(p+1)\times (p+1)$ matrix (including $\hat{G}_{00},
{\cal F}_{0i}$ as matrix elements).
Up to D-6 brane the polynomials $P_{(p)}$ of R-R potentials for
$D_{p}$ branes are (here we omit the hat $\wedge$ until equation(14))
\begin{equation}
\begin{array}{rl} 
D1 & 2\pi\alpha' F_{0i}C,\\[5mm]
D2 & 2\pi\alpha'\epsilon^{ij}F_{0i}C_{j},\\[5mm]
D3 & 2\pi\alpha'\epsilon^{ijk}F_{0i}
(\frac{1}{2}C_{jk}+\frac{1}{2}{\cal F}_{jk}C),\\[5mm]
D4 & 2\pi\alpha'\epsilon^{ijkl}F_{0i}
(\frac{1}{3!}C_{jkl}+\frac{1}{2}{\cal F}_{jk}C_{l})= 2\pi \alpha'
F_{0i} (^{\star}C^{i}_{(3)}+ ^{\star}{\cal F}^{ij}C_{(1)j}),\\[5mm]
D5 & 2\pi\alpha'\epsilon^{ijklm}F_{0i}
(\frac{1}{4!}C_{jklm}+\frac{1}{4}{\cal F}_{jk}C_{lm}+\frac{1}{8}{\cal
F}_{jk} {\cal F}_{lm}C),\\[5mm]
D6 & 2\pi\alpha'\epsilon^{ijklmn}F_{0i}
(\frac{1}{5!}C_{jklmn}+\frac{1}{12}{\cal F}_{jk}C_{lmn}+\frac{1}{8}{\cal 
F}_{jk} {\cal F}_{lm}C_{n}).
\end{array}
\end{equation}
Even $p's$ are part of IIA string while odd $p's$ are part of IIB string.
In the above result we have used the same assumption
$ g_{0i}=B_{0i}=0 $ and $\theta_{0i}=0$ as reference [4,10]. Hence we have
also $G_{0i}=
\Phi_{0i}=0$. We also omit the  time component of R-R potentials.
Their behavior under the Morita transformation can be obtained similarly, 
here we only consider R-R potentials with indeces in directions on a
torus $T^{p}$.

The interpolating formula proposed by Seiberg and Witten is [4],
\begin{equation}
 \frac{1}{G+2\pi{\alpha'}\Phi}=-\frac{\theta}{2\pi{\alpha'}}
+\frac{1}{g+2\pi{\alpha'}B}.
\end{equation}
Introduce $E= \frac {r^{2}(g+2\pi{\alpha'}B)}{{\alpha'}}$
and $\Theta =\frac{\theta}{2\pi r^{2}}$, in which $2\pi r$ is the period
for the D-p brane compactified on a torus $ T^{p}$ [4]. Now the T-duality
SO(p,p;Z) transformations are chosen as 
\begin{equation}
 E'= (aE+b)(cE+d)^{-1},
\end{equation}
and
\begin{equation}
 {\Theta'} =(c+d\Theta)(a+b\Theta)^{-1},
\end{equation}
where 
\begin{equation} 
T= \left(\begin{array}{cc}
a & b\\ c & d \end{array}\right) \in SO(p,p;Z)
\end{equation}
satisfying 
\begin{equation} 
T^{t}\left (\begin{array}{cc}
0 & 1\\ 1 & 0 \end{array}\right )T
=\left (\begin{array}{cc}
0 & 1\\ 1 & 0 \end{array}\right ).
\end{equation}
By using the following Morita transformation rules [4,10]:
\begin{eqnarray}
G'_{s}&=& \sqrt{detA} G_s,\quad\quad
G' = A G A^{t},\quad \quad A \equiv a + b\Theta ,\nonumber\\
{F'}+\Phi' &=& A ( F+\Phi) A^{t}\quad,\quad  {F'}_{0i}
=(F A^{t})_{0i}, 
\end{eqnarray}
and [9]
\begin{eqnarray}
STr'_{\theta'} &=& \frac{1}{\sqrt{\det A}}STr_{\theta}.
\end{eqnarray}
Ryang proved the invariance of $ {\cal L}_{DBI}$. 
He also discussed the Morita transformation law of R-R potentials
 in a complicated way.

However, if we consider the interpolation also in the $\hat {{\cal
L}}_{wz}$ term, we will get much  simpler Morita transformation rules for
R-R potentials. Concretly assume 
\begin{equation}
 \hat {{\cal L}}_{wz}=STr_{\theta}(e^{2\pi
\alpha'(\hat{F}+\Phi)}
\sum C_{(n)}),
\end{equation}
then we have
\begin{eqnarray}
C'&=& \frac{1}{\sqrt{\det A}}C,\;\;\; 
C'_{i}= \frac{1}{\sqrt{\det A}} A_{i}^{a}C_{a},\;\;\;
 C'_{ij}= \frac{1}{\sqrt{\det A}} (A C A^{t})_{ij},\nonumber\\
 C'_{ijk}&=& \frac{1}{\sqrt{\det A}} (A^{a}_{[i}
A^{b}_{j}A^{c}_{k]})C_{abc},\;\;\; 
C'_{ijkl}= \frac{1}{\sqrt{\det A}} (A^{a}_{[i}
A^{b}_{j}A^{c}_{k}A^{d}_{l]})C_{abcd},\nonumber\\
{C'}_{ijklm}&=& \frac{1}{\sqrt{\det A}} (A^{a}_{[i}
A^{b}_{j}A^{c}_{k}A^{d}_{l}A^{e}_{m]})C_{abcde}.
\end{eqnarray}

Similar to Yang-Mills fields, when the $\theta$ varies, the R-R
potentials should change as following to guarantee the correct
interpolation:\\
In IIA case,
\begin{eqnarray}
\delta \hat{C}(\theta)&=& -\hat{F}^{-1}_{0i}\delta
\hat{F}_{0i}\hat{C},\nonumber\\
\delta \hat{C}_{jk}(\theta)&=& -\delta(\hat{F}+\Phi)_{jk}\hat{C}
-\hat{F}^{-1}_{0i}\delta \hat{F}_{0i}\hat{C}_{jk},\nonumber\\
\delta \hat{C}_{jklm}(\theta)&=& -6\delta(\hat{F}+\Phi)_{jk}\hat{C}_{lm}
-\hat{F}^{-1}_{0i}\delta \hat{F}_{0i}\hat{C}_{jklm},
\end{eqnarray}
and in IIB case,
\begin{eqnarray}
\delta \hat{C}_{j}(\theta)&=& -\hat{F}^{-1}_{0i}\delta
\hat{F}_{0i}\hat{C}_{j},\nonumber\\
\delta \hat{C}_{jkl}(\theta)&=& -3\delta(\hat{F}+\Phi)_{jk}\hat{C}_{l}
-\hat{F}^{-1}_{0i}\delta \hat{F}_{0i}\hat{C}_{jkl},\nonumber\\
\delta \hat{C}_{jklmn}(\theta)&=&
-10\delta(\hat{F}+\Phi)_{jk}\hat{C}_{lmn}
-\hat{F}^{-1}_{0i}\delta \hat{F}_{0i}\hat{C}_{jklmn},
\end{eqnarray}
where  $\delta \hat{F}_{ij}(\theta)$ and $\delta \Phi_{ij}(\theta)$
follow from reference [4]:
\begin{eqnarray}
\delta \hat{F}_{ij}(\theta)&=&\frac{1}{4}\delta \theta^{kl}\{
2 \hat{F}_{ik}\star \hat{F}_{jl}+2 \hat{F}_{jl}\star \hat{F}_{ik}
-\hat{A}_{k} \star (\hat{D}_{l}\hat{F}_{ij}+\partial_{l}\hat{F}_{ij})
-(\hat{D}_{l}\hat{F}_{ij}+\partial_{l}\hat{F}_{ij}) \star \hat{A}_{k}\},
\nonumber \\
\delta \Phi_{ij}(\theta)&=&\delta \theta^{kl}(\hat{G}_{ik}\hat{G}_{lj}
+\Phi_{ik} \Phi_{lj}).
\end{eqnarray}

We can also find that the Morita equivalence in the Hamiltonian becomes
simpler when we consider the interpolation with modulus $\Phi$ in the
whole action. From our Lagrangian (4) and (5) it is easy to obtain 
\begin{equation}
{\cal H} = G_{00}^{1/2} STr_{\theta}\left\{
 \frac{-1}{G_s^2(2\pi)^{2p}\alpha'^{p+1}}det_{(p)}(G+{\cal F})
+\tilde{\epsilon}^{t}(G-{\cal F}G^{-1}{\cal F})
\tilde{\epsilon}\right\}^{1/2},
\end{equation}
in which $\det_{(p)}$ stands for the $p\times p$ matrix(without time
component) and 
\begin{equation}
 \tilde{\epsilon}^{i}=\epsilon^{i}-p^{i}_{(p)}(\hat{C},{\cal F}),
\quad\quad
p^{i}_{(p)}(\hat{C},{\cal F})\equiv \frac{\partial
P^{i}_{(p)}(\hat{C},{\cal F})}{2 \pi \alpha'\partial \hat{F}_{0i}},
\end{equation}
where $ \epsilon^{i}=\frac{\partial {L}}{\partial
\hat{F}_{0i}}$
are canonical momenta. It can be checked that $\det_{p}(G+{\cal F})$ is
equal to the trace term in reference [9] for $p=4$. Because of equation
(12) and equation (13)
the first part of Hamiltonian is obviouly invariant under Morita
transformation. The same is true if 
$\tilde{\epsilon}^{i}$ transform as follows
\begin{equation}
 \tilde{\epsilon'}^{i}=\sqrt{\det
A}[(A^{-1})^{t}\tilde{\epsilon}]^{i}.
\end{equation}
From the transformation of $\dot{A}_{i}=\hat{F}_{0i}$ in equation (12) and
note
the equation (13), we realize that 
\begin{equation}
\epsilon'^{i}=\sqrt{\det A}[(A^{-1})^{t}\epsilon]^{i}.
\end{equation} 
 To prove that
\begin{equation}
{p'}^i_{(p)}=\sqrt{\det A} [(A^{-1})^{t}p_{(p)}]^{i}.
\end{equation}
we can choose $D=4$ as an example. From equation (14)
we have 
\begin{equation}
 ^\star\hat{C'}_{(3)}^i \equiv
\frac{1}{3!}\epsilon^{ijkl} \hat{C'}_{jkl}=\sqrt{\det A}
[(A^{-1})^{t}\;^{\star}\hat{C}_{(3)})]^{i},
\end{equation}
and 
\begin{equation}
 ^{\star}({\hat{F}}+ \Phi)'^{ij}\equiv
\frac{1}{2}\epsilon^{ijkl}
({\hat{F}}+ \Phi)'_{kl}=(\det A)((A^{-1})^{t}\;^{\star}(\hat{F}+
\Phi)A^{-1})^{ij}.
\end{equation}
At last we get 
\begin{equation}
\begin{array}{rcl}
p'_{(4)}&=& ^{\star}\hat{C'}_{(3)}+\;^{\star}(\hat{F}+
\Phi)'\hat{C'}_{(1)}=\sqrt{\det A}
(A^{-1})^{t}(\;^{\star}\hat{C}_{(3)} +(\hat{F}+ \Phi)\hat{C}_{(1)})\\
&= & \sqrt{\det A}(A^{-1})^{t}\;p_{(4)}.
\end{array}
\end{equation}

Noncommutative parameter $\Theta(\theta)$ can be interpolated to zero, 
and $\Theta(\theta)$ can also be reduced to zero through Morita
transformation. But the latter case (different from the former) is
restricted
to the rational value. Interpolating parameter must be very small, while
the Morita parameter may not have this restriction. It was argued that
modulus $\Phi$ appears as a magnetic backgroud. Under this magnetic
background the Morita transformation becomes simpler. That is due to  the
Morita symmetry of $\Phi$ compensating  the inhomogeneous change of gauge
field. More property of $\Phi$ can be found from Reference [6,11].
\vspace{5mm}

{\hspace{5mm} {\Large  \bf Acknowledgement}
\vspace{5mm}

The authors are benefited from the String and Noncommutative geometry 
Workshop held at Tsinghua University in June 2000.
They would like to thank Y.S. Wu and H. Liu for introducing
Ryang's paper[10] and valuable discussions.
They  would also like to thank Z.X. Yang for interesting discussions. This
work was partly supported by the National Natural Science Foundation of
China. 

\end{document}